\documentstyle[12pt]{article}
\newcommand{\doublespacing}{\let\CS=\@currsize\renewcommand{\baselinesstrech}
{2.0}\tiny\CS}

\begin{document}

\title{Quantum dot with spin-orbit interaction in noncommutative phase space and analog Landau levels}
\author{B.K.Pal \thanks{E-mail: barunpal1985@rediffmail.com}, B. Roy\thanks{E-mail :
barnana@isical.ac.in}, B. Basu \thanks{E-mail:
banasri@isical.ac.in ~~Fax:91+(033)2577-3026} \\ Physics \& Applied
Mathematics Unit\\ Indian Statistical Institute \\ Calcutta  700108\\ India}

\maketitle

\vspace*{1.5cm}
\begin{abstract}
We have studied a quantum dot with Rashba spin orbit interaction
on a plane where the position and momentum coordinates are
considered to be noncommutative. The energy spectrum of the system
is found to be equivalent to that of a quantum dot  with Rashba
spin-orbit interaction in a magnetic field under certain
conditions.
\end{abstract}
\section{Introduction}
In recent years considerable attention has been paid to the
investigation of different physical problems in noncommutative
spaces [1]. The reasons for the emergence of this interest were
the predictions of String theory [2] in the low-energy limit
which, along with the Brane-world scenario [3], led to the fact
the space-time could be noncommutative. Later intensive
investigation of the field theory on noncommutative spaces was
prompted by M-theory [4] and the matrix formulation of the quantum
Hall effect [5]. It has been shown that noncommutative spaces may
arise as the gravitation quantum efeect and  and may serve as a
possible way for the regularization of quantum field theories [6].
By considering the low-energy limit of one-particle sector of
field theory on noncommutative space one arrives at what is called
noncommutative quantum mechanics. Although the noncommutative
scale is expected to be pretty small, perhaps below the planck
scale [7], people are looking for phenomenological consequences of
the noncommutative geometry in low energy quantum mechanical
regime. Many interesting quantum mechanical problems on
noncommutative spaces have been investigated and the effect of
noncommutativity on the observables has been analyzed [8]. 

On the
other hand, in recent years, the physics concerning the spin of
electrons has become a widely studied subject in the context of
semiconductor physics. The fundamental physics involved and the
possibility of future technological applications motivate this
research . In particular, the role of the spin in low dimensional
semiconductor structures like quantum dots, is especially
attractive because of the ability to control the relevant
parameters of the nanostructures available at present. Thus it is
possible to have knowledge of the interactions governing the spin
dynamics within these structures. Among the interactions
concerning the spin, a relevant intrinsic interaction in
non-magnetic semiconductors is the spin-orbit coupling, which has
been extensively studied in the context of quantum dots [9]. This
interaction stems from the relativistic correction to the electron
motion. The electric fields present in the semiconductor are felt
by the electrons in their intrinsic reference frame as a
spin-dependent magnetic field whose intensity depends on the
momentum of the particle. Depending on the origin of the electric
field there are two types of spin-orbit coupling: the lack of
inversion symmetry in a bulk semiconductor gives rise to the
so-called Dresselhaus term [10], whereas the asymmetry of the
potential confining the two-dimensional electron gas of a
heterostructure is responsible for Bychkov-Rashba interaction
[11]. The other spin-dependent interaction is the Zeeman
interaction arising from the direct coupling of the intrinsic
magnetic moment associated with the spin with an applied magnetic
field. In this regard, the analytical solutions for quantum dots with Rashba spin-orbit interaction for diferent confinement potential is also studied \cite{10,11,12}.

In this note, we shall consider a quantum dot with Rashba spin-orbit coupling in a space where both position and momentum coordinates are noncommutative. The reason for taking the momentum noncommutativity comes from the fact that in quantum mechanics, the generalized momentum components are noncommutative. Moreover, it has been shown recently \cite{jiang} that in order to keep the Bose-Einstein statistics for identical particles intact at the noncommutative level, one should consider both the space-space  and momentum-momentum noncommutativity.

 This note is organized as follows: In Section 2 we consider a quantum dot with harmonic confinement and Rashba spin-orbit interaction in a two dimensional space where both the position and momentum are noncommutative. The energy spectrum of the system is found and its degeneracy is discussed for specific noncommutative parameter ranges. In Section 3 the model of Section 2 is compared with a system consisting of a quantum dot with Rashba spin-orbit interaction  in a magnetic field and a relation between the noncommutative parameters $\theta $ and $\bar{\theta}$ with the magnetic field and Rashba spin-orbit interaction coupling strength $\lambda_R$ is established. Finally, the concluding remarks are given in Section 4.

% In the literature, the conditions under which the Landau problem and the isotropic oscillator in noncommuative space are equivalent have been explored \cite{gamboa,pr}. This motivates us to consider a quantum dot in two dimensions with Rashba spin-orbit coupling in a noncommutative space and   quantum dot in magnetic field and find the condition of equivalence.

Thorughout our calculations we have used the natural unit system~($\hbar=c=1$) and we have retained the terms upto the order of $\lambda_R^2$.

\section{Quantum dot with spin-orbit interaction in noncommutative space}

In the presence of Rashba spin orbit interaction, the Hamiltonian of a single electron state of a quantum dot with harmonic confinement is given by,
\begin{equation}
H=-\frac{1}{2m}~~(\frac{\partial^2}{\partial
x^2}+\frac{\partial^2}{\partial y^2})+\frac{m}{2}
\omega_{0}^2(x^2+y^2)+i~\lambda_{R}(\sigma_{x}\frac{\partial}{\partial
y}~-\sigma_{y}\frac{\partial}{\partial x})
\end{equation}
where $\lambda_R$ is the Rashba  spin-orbit coupling parameter.

In non-commutative spacetime the ordinary product is replaced by a
star product of the form
\begin{equation}\label{moyal}
\psi(x)\star \phi(x)=\exp\{i\theta_{\mu\nu}\partial_x^{\mu}\partial_y^{\nu}\} \psi(x)\phi(x)|_{x=y}
\end{equation}
where $\theta_{\mu\nu}$ is an antisymmetric constant matrix. As
$\theta_{0i}\neq 0$ leads to a non-unitary theory \cite{gomis},
only the elements $\theta_{ij}, i,j=1,2,3$ are allowed to be
nonvanishing. In order not to spoil the isotropy os space it is
mandatory to choose $\theta_{ij}$ proportional to the constant
antisymmetric matrix $\epsilon$,
$\theta_{ij}=\theta\varepsilon_{ij}$, where $\theta$, the
noncommutativity parameter, is a constant with dimension of
$(length)^2$ and
\begin{equation}
\epsilon=\left( \begin{array}{ccc}
0 & -1& 1 \\
1 & 0 & -1 \\
-1 & 1 &0
\end{array}\right)
\end{equation}
  This leads to the Moyal comutation relations between the spatial and momentum coordinates as
 %To go over to the noncommutative space we replace the commutative quantities by the noncommuative ones and
 % and denote them by a bar .\\
%\\ as we consider both the position and the momentum space to be
%non-commutative, the non-commutative algebra in phase space can be
%written as,
\begin{equation}
\begin{array}{lcl}
     [{\bar{x_i}},{\bar{x_j}}]&= & 2i\varepsilon_{ij}\theta  \\

     [{\bar{p_i}},{\bar{p_j}}]&= & 2i \varepsilon_{ij}{\bar{\theta}} \\

     [{\bar{x_i}},{\bar{p_j}}]&= & i{\delta_{ij}}[1+\theta{\bar{\theta}}]
     \end{array}
     \end{equation}
     %\end{document}
     Here $\theta$ is the non commutative parameter for the co-ordinate
     space and $\bar{\theta}$ is the non commutative parameter for the
     momentum space.
    % $\varepsilon_{11}=\varepsilon_{22}=0,~~~\varepsilon_{12}=-\varepsilon_{21}=1$.
     The non-commutative  Hamiltonian is then given by,
     \begin{equation}
     {\bar{H}}=\frac{1}{2m}{\bar{{\bf {p}}}}^2 +\frac{1}{2}m\omega_0^2 {\bar{\bf {r}}}^2+\lambda_R(\sigma_x {\bar{p_y}}-\sigma_y{\bar{p_x}})
     \end{equation}

 Now the star product (\ref{moyal}) can be replaced by a shift called Bopp's shift \cite{curt} so that
the non-commutative operators and commutative operators are
     related via the following relation:
     $$ \bar{x}=x-\theta p_y~~,~~\bar{y}=y+\theta p_x$$
           $$\bar{p_x}=p_x+\bar{\theta}y~~,~~\bar{p_y}=p_y-\bar{\theta}x.$$
           Hence,  the above non-commutative Hamiltonian in the
           commutative space  can be written as:
          \begin{equation}\label{ham1} H_{NC}=A(p_x^2+p_y^2)+B(x^2+y^2)+C(yp_x-xp_y)+\lambda_R[(p_y\sigma_x-\sigma_yp_x)-\bar{\theta}(x\sigma_x+y\sigma_y)]=H_0+H_{spin}
 \end{equation}
          where,
 $$ ~~~A=(\frac{1}{2m}+\frac{m\omega_0^2\theta^2}{2}),$$
   $$ B=(\frac{\bar{\theta^2}}{2m}+\frac{m\omega_0^2}{2}),$$
$$C=(\frac{\bar{\theta}}{m}+\theta m \omega_0^2),$$
By means of an unitary transformation \cite{alneir, rod} the above
Hamiltonian can be diagonalised in spin space upto second order in
the spin-orbit parameter $\lambda_R$. In order to neglect higher
order terms introduced by the unitary transformation it is
required that the energy scale correponding to the spin orbit
interaction must be much smaller than that corresponding to the
spin independent Hamiltonian $H_0>>H_{spin}$. The  Hamiltonian
(\ref{ham1}) in the transformed reference frame is given by
\begin{equation}\label{ham2}
H^\prime=U^\dag {H}_{NC}U
\end{equation}
where,
\begin{equation}
\begin{array}{lcl}
U &=& e^{-i\beta\hat{O}}\\
\hat{O}&= &(p_x-\frac{y}{\theta})\sigma_x+(p_y-\frac{x}{\theta})\sigma_y\\
\beta &=& \frac{\lambda_R}{~~\frac{\bar{\theta}}{m}-\frac{1}{m\theta}}
\end{array}
\end{equation}

We expand the transformed Hamiltonian (\ref{ham2})upto the 2nd
order in $\beta$ using the Baker-Hausdroff Lemma as:

\begin{equation}\label{ham3}
\begin{array}{lcl}
H^\prime &\approx& {H}_{NC}
+i\beta[\hat{O},{H}_{NC}]-\frac{\beta^2}{2}[\hat{O},[\hat{O},{H}_{NC}]]
\\\\
 &=&\frac{\bf p^2}{2M_{\theta{\bar{\theta}}}}+ \frac{M_ {\theta{\bar{\theta}}}\Omega^2_{\theta{\bar{\theta}}}\bf
r^2}{2}-S_{\theta{\bar{\theta}}} L_z -\lambda_R^2 m
\end{array}
\end{equation}
 where
\begin{equation}
\begin{array}{lcl}
 M_{\theta{\bar{\theta}}} &= &[~~\frac{1}{m}+m\omega_0^2 \theta^2+\frac{2\lambda_R^2
m}{(\bar{\theta}-\frac{1}{\theta})}~~\sigma_z~~]^{-1}
\\\\
\Omega^2_{\theta{\bar{\theta}}}&=& [~~\frac{1}{m}+m \omega_0^2
\theta^2+\frac{2 \lambda_R^2
m}{(\bar{\theta}-\frac{1}{\theta})}~~\sigma_z~~]~[~~\frac{\bar{\theta^2}}{m}+
m\omega_0^2-\frac{2 \lambda_R^2
m\frac{\bar{\theta}}{\theta}}{(\bar{\theta}-\frac{1}{\theta})}~~\sigma_z~~]\\\\
S_{\theta{\bar{\theta}}}&= &[~~\frac{\bar{\theta}}{m}+m \omega_0^2
\theta + \lambda_R^2 m~\sigma_{z}~~]\\ \\
L_z &=&(xp_y-yp_x)
\end{array}
\end{equation}

 It is noted that $[H^\prime,\sigma_z]=0 $, and so we can use the spin on the $z$-direction to characterize the states.
The Hamiltonian (\ref{ham3}) takes the form
\begin{equation}
H^\prime =\left(%
\begin{array}{cc}
  H_{NC}^\uparrow & 0 \\
  0 & H_{NC}^\downarrow \\
\end{array}%
\right)
\end{equation}
where
 \begin{equation}\label{ham4}
\begin{array}{lcl}
H^{\uparrow, \downarrow}_ {NC}= \frac{\bf
p^2}{2}[\frac{1}{m}+m\omega^2_{0}\theta^2\pm \frac{2 \lambda_R^2
m}{(\bar{\theta}-\frac{1}{\theta})}]+\frac{\bf
r^2}{2}[~~\frac{\bar{\theta^2}}{m}+ m\omega_0^2\mp \frac{2
\lambda_R^2
m\frac{\bar{\theta}}{\theta}}{(\bar{\theta}-\frac{1}{\theta})}]-L_{z}
[~~\frac{\bar{\theta}}{m}+m \omega_0^2 \theta \pm \lambda_R^2
m~~]-\lambda_R^2 m
\end{array}
\end{equation}\\
The term $\lambda_R^2 m$ is a constant for a particular system
with a fixed spin orbit interaction. Then the Hamiltonians
$H_{\theta{\bar{\theta}}}^{\uparrow, \downarrow}=H^{\uparrow,
\downarrow}_ {NC}+\lambda_R^2 m$ can be solved exactly.  It is
possible to write any of the Hamiltonians $
H_{\theta{\bar{\theta}}}^{\uparrow,\downarrow}$  as a sum of two
separate one dimensional harmonic oscillators with frequencies
${\Omega^+}_{\theta{\bar{\theta}}}$ and
${\Omega^-}_{\theta{\bar{\theta}}}$ \cite{15,16,17}. As
$H^{\uparrow}_{\theta{\bar{\theta}}}$ and
$H^\downarrow_{\theta{\bar{\theta}}} $ are similar, we shall
discuss only the case for $H_{\theta{\bar{\theta}}}^{\uparrow}$.
For convenience, we are dropping the $\uparrow$ sign.
\begin{equation}
H_{\theta{\bar{\theta}}}
%^{(\uparrow,\downarrow)}
=\Omega_{\theta{\bar{\theta}}}^{+}
%(\uparrow,\downarrow)~
[~a_{\theta{\bar{\theta}}}^{\dag}
%(\uparrow,\downarrow)~
a_{\theta{\bar{\theta}}}
%(\uparrow,\downarrow)
+\frac{1}{2}]+\Omega_{\theta{\bar{\theta}}}^{-}
%(\uparrow,\downarrow)~
[~b_{\theta{\bar{\theta}}}^{\dag}
%(\uparrow,\downarrow)~
b_{\theta{\bar{\theta}}}
%(\uparrow,\downarrow)
+\frac{1}{2}]
\end{equation}
The explicit forms of the frequencies are given by
\begin{equation}
\Omega_{\theta{\bar{\theta}}}^{+}
%(\uparrow)
=\sqrt{\Omega_{\theta{\bar{\theta}}}^2
%(\uparrow)
-S_{\theta{\bar{\theta}}}^2}
%(\uparrow)}
+S_{\theta{\bar{\theta}}},
%(\uparrow),
\end{equation}

\begin{equation}
\Omega_{\theta{\bar{\theta}}}^{-}
%(\uparrow)
=\sqrt{\Omega_{\theta{\bar{\theta}}}^2
%(\uparrow)
-S_{\theta{\bar{\theta}}}^2}
%(\uparrow)}
-S_{\theta{\bar{\theta}}}
%(\uparrow)
\end{equation}
where $\Omega_{\theta{\bar{\theta}}}^{+}\neq \Omega_{\theta{\bar{\theta}}}^{-}$.
Also, the anihilation operators are given by
%$a_{\theta{\bar{\theta}}}(\downarrow) , ~~ b_{\theta{\bar{\theta}}}(\downarrow)~ and~~ \Omega_{\theta{\bar{\theta}}}^{+}(\downarrow)~~,\Omega_{\theta{\bar{\theta}}}^{-}(\downarrow)
$a_{\theta{\bar{\theta}}}, b_{\theta{\bar{\theta}}}$ are given by
\begin{equation}
a_{\theta{\bar{\theta}}}
%(\uparrow)
=\frac{1}{2\sqrt{M_{\theta{\bar{\theta}}}
%(\uparrow)
\Omega_{\theta{\bar{\theta}}}^{+}}}
%(\uparrow)}}
~[~(p_x+ip_y)-iM_{\theta{\bar{\theta}}}
%(\uparrow)
\Omega_{\theta{\bar{\theta}}}^{+}
%(\uparrow)
(x+iy)~]
\end{equation}
\begin{equation} b_{\theta{\bar{\theta}}}=\frac{1}{2\sqrt{M_{\theta{\bar{\theta}}}\Omega_{\theta{\bar{\theta}}}^{-}}}~[~(p_x-ip_y)-iM_{\theta{\bar{\theta}}}\Omega_{\theta{\bar{\theta}}}^{-}(x-iy)~]
\end{equation}
The commutation relations between the creation operators  $a_{\theta{\bar{\theta}}}^\dag, b_{\theta{\bar{\theta}}}^\dag$ and the annihilation operators $a_{\theta{\bar{\theta}}}, b_{\theta{\bar{\theta}}}$, are
\begin{equation}
[a_{\theta{\bar{\theta}}},a_{\theta{\bar{\theta}}}^{\dag}]=[b_{\theta{\bar{\theta}}}
%(\uparrow)
,b_{\theta{\bar{\theta}}}^{\dag}]
%(\uparrow)]
=1
\end{equation}
with all other commutations being zero.
The number operators
$N_{\theta{\bar{\theta}}}^{+}=a_{\theta{\bar{\theta}}}^{\dag}~a_{\theta{\bar{\theta}}}$~
and~$N_{\theta{\bar{\theta}}}^{-}=b_{\theta{\bar{\theta}}}^{\dag}~b_{\theta{\bar{\theta}}}$~satisfy
the following eigenvalue eqation
\begin{equation}
N_{\theta{\bar{\theta}}}^{\pm}~|~~n^{+},~n^{-}~~\rangle~=n^{\pm}~|~~n^{+},~n^{-}~~\rangle
\end{equation}
with $n^{\pm}=0,1,2...........$.\\
The energy eigenvalues of the Hamiltonian~$H_{nc}$ given in eqn.(13) are
\begin{equation}
E_{\theta{\bar{\theta}}}=\Omega_{\theta{\bar{\theta}}}^{+}~(~n^{+}+\frac{1}{2})+\Omega_{\theta{\bar{\theta}}}^{-}~(~n^{-}+\frac{1}{2})
\end{equation}
which is non-degenerate due to the presence of noncommutativity.

Now we can study some special cases concerning different relations between the two noncommutative parameters:\\
(i)  If $\bar{\theta}=-m^2\omega_0^2~\theta-\lambda_R^2~m^2$, then $S_{\theta{\bar\theta}}=0$, and the energy eigenvalue is given by
\begin{equation}
E_{\theta}=(n^+~+~n^-~+1)\Omega_\theta
\end{equation}
where
\begin{equation}
\Omega^+ =\Omega^-=\Omega_\theta=\omega_0\sqrt{(1+m^2\omega_0^2\theta^2)^2+2m^2\theta\lambda_R^2(m^2\omega_0^2\theta^2-1)}
\end{equation}
It can be seen from eqn.(22) when $\lambda_R = 0$, the Hamiltonian is of an isotropic harmonic oscillator and our result coincides with that of ref.\cite{pr}.
\\

(ii)For, ${\bar\theta}=\frac{1}{\theta}+2\lambda_R~ m~\frac{1}{\sqrt{\theta}}$, $\Omega_{\theta{\bar\theta}}=S_{\theta{\bar\theta}}$, then
\begin{equation}
E_{\theta}=(n^+~-~n^-)\Omega_\theta
\end{equation}
where
\begin{equation}
\Omega^+ =-\Omega^-=\Omega_\theta=\frac{1}{m\theta}+\frac{2\lambda_R}{\sqrt{\theta}}+m\omega_0^2\theta+\lambda_R^2m
\end{equation}
The energy spectrum (23) is infinitely degenerate due to the constraint
${\bar\theta}=\frac{1}{\theta}+2\lambda_R~ m~\frac{1}{\sqrt{\theta}}$ which reduces to the constraint $\theta{\bar\theta}=1$ given in ref.\cite{pr} when $\lambda_R = 0$.\\

(iii) The properties of the spectrum (12) depend on the ratio $\frac{\Omega_{\theta{\bar{\theta}}}^{+}}{\Omega_{\theta{\bar{\theta}}}^{-}}$. For irrational $\frac{\Omega_{\theta{\bar{\theta}}}^{+}}{\Omega_{\theta{\bar{\theta}}}^{-}}$ the spectrum is non-degenerate while rational $\frac{\Omega_{\theta{\bar{\theta}}}^{+}}{\Omega_{\theta{\bar{\theta}}}^{-}}$ leads to degeneracy \cite{15}. Let us consider the latter case. We assume that $\frac{\Omega_{\theta{\bar{\theta}}}^{+}}{\Omega_{\theta{\bar{\theta}}}^{-}} = \frac{a^-}{a^+}$ where $a^-$,$a^+$ are relatively prime numbers. Let us put $\Omega_{\theta,\bar {\theta}}^+ = a^- \sigma$ and $\Omega_{\theta,\bar {\theta}}^- = a^+ \sigma$ where $\sigma$ is a constant. Then equation (12) gives
\begin{equation}
E_{\theta, \bar {\theta}} = \sigma (a^- n^+~+ a^+ n^-) + \sigma \left (\frac{a^- + a^+}{2}\right )
\end{equation}
The last term on the right hand side of equation (25) is an overall constant. So it follows that the spectrum (25) is degenerate, the degree of degeneracy being equal to the number of natural solutions $n^{\pm}$ to the equation $a^- n^+~+ a^+ n^- = \rm constant$.

%where .
%$$The ground state of the Hamiltonian corresponds
%to~$n_{\theta{\bar{\theta}}}^{\pm}(\uparrow)=0$~and this energy is
%given by,
%$$E_{\theta{\bar{\theta}}}^{'}(\uparrow)=\omega_0[1+\bar{\theta}^2\theta^2+2\lambda_R^2m^2(1+\theta)]^{\frac{1}{2}}$$\\

\section{Quantum dot with spin-orbit interaction in a magnetic field: Conditions for equivalence}

In the literature, an inverse relation between the noncommutative parameter and magnetic field has been obtained by comparing term by term of the Hamiltonian of the Landau problem and the Hamiltonian of an oscillator in noncommutative coordinate space. \cite{gamboa}.
%In particular, if the potential in noncommuative quantum mechanics is chosen as $V=\Omega \cal{N}$ where $\Omega$ is a constant and
%\begin{equation}
%\begin{array}{lcl}
%{\hat{\cal{N}}}&=&\frac{\theta^2}{4}p_x^2+x^2+\frac{\theta^2}{4}p_y^2 +y^2-\theta L_z\\
%&=& {\hat{H}}_{HO}-\theta {\hat{L}}_z
%\end{array}\end{equation}
 %${\hat{H}}_{HO}$ being the Hamiltonian for a two dimensional harmonic oscillator with mass $\frac{2}{\theta^2}$, frequency $\omega=\theta$ and $L_z$ being the z-component of the angular momentum defined as ${\hat{L}}_z=xp_y-yp_x$, the Landau problem and the noncommutative quantum mechanics are equivalent theories in the lowest Landau level \cite{gamboa}.
  Moreover, considering the isotropic oscillator on a plane and taking both the position and momentum coordinates as noncommuative it was shown in \cite{pr} that there is a relation between the two noncommuative parameters with the magnetic field of the Landau problem, given by ${\bar\theta}=\frac{1}{\theta}=\frac{B}{2}$. This has motivated us to find whether there is a  equivalence between  the Hamiltonians representing a quantum dot with spin orbit interaction in noncommutative space with that of a quantum dot with spin orbit interaction in a magnetic field.
In ref. \cite{rod} a partial diagonalization of the Hamiltonian representing a quantum dot with spin orbit interaction and Zeeman energy on an equal footing was derived analytically. Since we shall require the above mentioned results of
ref. \cite{rod}, we reproduce it here briefly.

The Hamiltonian representing a quantum dot with spin orbit interaction and Zeeman energy is given by,
\begin{equation}
 H_B=H_0 + H_R + H_z
 \end{equation}
 where,
 \begin{equation}
 \begin{array}{lcl}
 H_0&=&\frac{1}{2m}(P_x^2+P_y^2)+\frac{1}{2} m
\omega_0^2(x^2+y^2)\\ \\
 H_R &=&\lambda_R(P_y\sigma_x-P_x\sigma_y)\\
H_z &=&\frac{\varepsilon_z}{2} \sigma_z
\end{array}
\end{equation}
Here, ${\bf P}={\bf p}+\frac{e}{c}{\bf A}~~,~~{\bf A}=\frac{B}{2}(-y,x,0)$
 This Hamiltonian can be diagonalized in spin space upto second order in the spin-orbit and the Zeeman parameters with the help of an unitary transformation\cite{rod}. It is to be mentioned that the energy scale corresponding to the spin-orbit and Zeeman interactions must be much smaller than that corresponding to the spin independent Hamiltonian ($H_0>>H_R, H_z$), in order to neglect higher order terms introduced by the unitary transformation.

The transformed Hamiltonian is given by ,
\begin{equation}
H_B^{'}=U^{\dagger}H_BU~~,U=e^{-i\beta\hat{O}}
\end{equation}
where,
\begin{equation}
{\hat{O}}=(P_x- \frac{m \omega_0^2}{\varepsilon_z}y)\sigma_x +(P_y
+ \frac{m\omega_0^2}{\varepsilon_z}x) \sigma_y
\end{equation}
\begin{equation}
 \beta=-\frac{\lambda_R
\varepsilon_z}{\omega_0^2+\omega_c\varepsilon_z-\varepsilon_z^2}~,~~~~~~\omega_c=\frac{eB}{mc}
\end{equation}
%\begin{equation}
We expand the transformed Hamiltonian upto the 2nd order of~~$
\beta$~using Baker Hausdrauff lemma as,
\begin{equation}
\begin{array}{lcl}
\displaystyle H_B^{'} &=& H_B + i\beta[\hat{O},H_B] -
\frac{\beta^2}{2}[\hat{O},[\hat{O},H_B]] \nonumber \\\\
&=& \frac{1}{2M_1}(p_x^2+p_y^2)+\frac{1}{2}M_1
\Omega_1^2(x^2+y^2)-S_1(xp_y-yp_x)+\frac{\varepsilon_z}{2}~\sigma_z-K
\end{array}
\end{equation}
where,
\begin{equation}
\frac{1}{ M_1}=\frac{1}{m} \left[1-\frac{2 \lambda_R^2 m
 \varepsilon_z}{\omega_0^2+\omega_c\varepsilon_z-\varepsilon_z^2}~\sigma_z \right]
\end{equation}
\begin{equation}
\Omega_1^2=M_1^{-1}[m \omega_0^2+\frac{e^2 B^2}{4
c^2}-\frac{\lambda_R^2 \omega_0^2 m}{\omega_0^2
+\omega_c\varepsilon_z -\varepsilon_z^2}\frac{eB}{c} \sigma_z]
\end{equation}
\begin{equation}
K=\lambda_R^2 m[\frac{\omega_0^2+\omega_c
\varepsilon_z}{\omega_0^2+\omega_c \varepsilon_z
-\varepsilon_z^2}]
\end{equation}
\begin{equation}
S_1=\frac{\lambda_R^2 \omega_0^2 m}{\omega_0^2
+\omega_c\varepsilon_z -\varepsilon_z^2} \sigma_z-\frac{eB}{2c
M_1}
\end{equation}
 Now we define,
 \begin{equation}
\begin{array}{lcl}\displaystyle
~H_{1B}^{'}&=&H_B^{'}+K-\frac{\varepsilon_z}{2}~\sigma_z  \\\\
\displaystyle~~~~~&=&\frac{1}{2M_1}(p_x^2+p_y^2)+\frac{1}{2} M_1
\Omega_1^2(x^2+y^2)-S_1(xp_y-yp_x) \nonumber \\\\
\displaystyle~~~~~&=&\Omega_1^+(a_1^\dagger
a_1+\frac{1}{2})+\Omega_1^-(b_1^\dagger
b_1+\frac{1}{2})
\end{array}
\end{equation}
The Hamiltonian $H_{1B}^{'}$ for the two spin eigenstates can be compared with the well known two dimensional harmonic oscillator problems.
Here, the corresponding frequencies are given by
\begin{equation}
\Omega_1^+=\sqrt{\Omega_1^2-S_1^2}+S_1~,~~~~~\Omega_1^{-1}=\sqrt{\Omega_1^2-S_1^2}-S_1
\end{equation}
and the annihilation operators are given by
\begin{equation}\begin{array}{lcl}
a_1 &=&\frac{1}{2\sqrt{ M_1 \Omega_1^+}}~[(p_x+ip_y)-iM_1 \Omega_1^+(x+iy)],\\b_1 &=&\frac{1}{2\sqrt{ M_1 \Omega_1^-}}~[(p_x-ip_y)-iM_1 \Omega_1^-(x-iy)]\end{array}\end{equation}
Correspondingly we can define the creation operators also and they satisfy the usual commutation relations.

Eqn.(36) can be written as
\begin{equation}
\begin{array}{lcl}\displaystyle ~~{H_{1B}^{'}}&=&\frac{
p^2}{2M_{1}}+\frac{M_{1} \Omega^2_{1}r^2}{2}-S_{1} L_z\\ \\
\displaystyle ~~~~~~~ &=& \Omega^+(a_1^\dagger a_1 + \frac{1}{2}) +
\Omega^-(b_1^\dagger b_1 +\frac{1}{2})
\end{array}
\end{equation}

The energy eigenvalues of this  Hamiltonian~$H_{1B}^\prime $ is given by
\begin{equation}
E_{1B}^\prime =\Omega_{1}^{+}~(~n^{+}+\frac{1}{2})+\Omega_{1}^{-}~(~n^{-}+\frac{1}{2})
\end{equation}
where $n^{\pm}=0,1,2,....$\\

The similarity of the energy spectrum corresponding to the Hamiltonians (9) and (26) can be exploited to find a relation between the noncommutative parameters and spin orbit coupling strength, and the magnetic field strength.
Comparing the like terms of the Hamiltonians (9) and (26) we get
\begin{equation}
\frac{2 \lambda_R^2
\varepsilon_z}{\omega_0^2+\omega_c\varepsilon_z-\varepsilon_z^2}+m
\omega_0^2\theta^2+\frac{2m\lambda_R^2}{\bar\theta-\frac{1}{\theta}}=0
\end{equation}

\begin{equation}
\frac{e^2B^2}{4m}\left(1-\frac{2\lambda_R^2 m
\varepsilon_z}{\omega_0^2+\omega_c \varepsilon_z
-\varepsilon_z^2}\right)+
m\omega_0^2-\frac{eBm\omega_0^2\lambda_R^2}{\omega_0^2+\omega_c
\varepsilon_z
-\varepsilon_z^2}=\bar\theta^2+m\omega_0^2-\frac{2m\lambda_R^2\frac{\bar\theta}{\theta}}{\bar\theta-\frac{1}{\theta}}
\end{equation}
\begin{equation}
\frac{m\omega_0^2\lambda_R^2}{\omega_0^2+\omega_c \varepsilon_z
-\varepsilon_z^2}-\frac{eB}{2m}(1-\frac{2m\lambda_R^2\varepsilon_z}{\omega_0^2+\omega_c
\varepsilon_z
-\varepsilon_z^2})=(\frac{\bar\theta}{m}+m\omega_0^2+\lambda_R^2m)
\end{equation}\\
\begin{equation}
\omega_0^2+\omega_c\epsilon_z-\epsilon_z^2=2\lambda_R^2 m~\epsilon_z
\end{equation}
The eqns.(37),(38), (39) and (40) are consistent if the folowing equation is satisfied
\begin{equation}
eB(\bar{\theta}+m^2\omega_0^2\theta+
\lambda_R^2~m^2)+{\bar{\theta}}^2
+\frac{{\bar\theta}}{\theta}(1+m^2\omega_0^2\theta^2)=0
\end{equation}
It is to be noted that if ${\bar\theta}=0$, then the equivalence condition (44) reduces to
\begin{equation}
\lambda_R^2=-\omega_0^2\theta
\end{equation}
which is independent of $B$. This result is consistent with that obtained in ref. \cite{16} which indicates that the origin of momentum noncommutativity is due to the magnetic field \cite{16}.
.
%We can make a comment on the equivalence of both the systems by comparing the energy eigenvalue eqns.(20) and (35).similar for spin down electrons

\section{Conclusion}
In this article, a quantum dot,in harmonic confinement, with Rashba spin -orbit interaction on a noncommutative phase space is considered which has not been reported in the literature so far our knowledge goes. We have obtained the energy spectrum which is nondegenerate. But the spectrum becomes degenerate when the noncommutative parameters $\theta$ and $\bar\theta$ satisfy certain constraints. Also we have compared our system with a quantum dot with Rashba spin-orbit interaction  in a magnetic field and derived the condition for  equivalence of the two systems when the spin -orbit interaction strength $\lambda_R$ is small. We intend to analyse the energy spectrum of a quantum dot in anisotropic confinemnet on a noncomutative plane in future.

\section{Acknowledgements} B.K.Pal acknowledges Council of  Scientific and Industrial  Research(CSIR), India for financial  support.

\section{References}
\begin{enumerate}
\item  M.R.Douglas and N.A.Nekrasov, Rev.Mod.Phys. {\bf 73} 977 (2001)\\R. Szabo Phys. Rep.
 378 207 (2003)\\
         A.P.Balachandran, T.R.Govindarajan, C.Molina and P.Teotonio-Sobrinho, JHEP {\bf 0410} 072 (2004)\\
\item  E.Witten, Nucl.Phys. {\bf B460} 33 (1996)\\
         M.R.Douglas and C.M.Hull, JHEP {\bf 02} 008 (1998)\\
         C.Chu and P.Ho, Nucl.Phys. {\bf B550} 151 (1999); ibid {\bf B568} 447 (2000)\\
         N.Seiberg and E.Witten, JHEP {\bf 09} 032 (1999)\\
\item  I.Antoniads, N.Arkani-Hamed, S.Dimopoulas and G.R.Dvali, Phys.Letts. {\bf B436} 257 (1998)\\
\item  A.Connes, M.R.Douglas and A.Schwarz, JHEP {\bf 02} 003 (1998)\\
  \item  B. Chakrabarty, S. Gangopadhyay and A. saha, Phys. Rev. D 70, 107707 (2004)\\
   O.F. Doyi and A. Jellal J. Math Phys. 43 4592 (2004)\\
 \item  H.S.Snyder, Phys.Rev. {\bf D71} 38 (1946)\\
         C.N.Yang, Phys.Rev. {\bf D72} 874 (1947)\\
  \item  F. Delduc et. al. J.Phys. Conf. Series {\bf 103} 012020 (2008)
  \item The literature is very extensive, some papers are:\\

  G.V.Dunne,J.Jackiw and
  C.Trugenberger, Phys.Rev.{\bf D41} 661 (1990).\\
  H.Falomir,J.Gamboa,M.Loewe,F.Mendez, Phys.Rev.{\bf D66} 045018 (2002)\\
  F.S.Bemfica,H.O.Girotti, J.Phys.{\bf A38} L539 (2005).\\
  C.Duval And P.A.Horvathy, Phys.Lett.{\bf B547} 306 (2002)\\
 C.Duval And P.A.Horvathy, J.Phys.{\bf A34} 10097 (2001),ibid,Phys.Lett. C.Duval And P.A.Horvathy,Phys.Lett.{\bf B479} 284 (2000)\\
D.V.Vassilevich,JHEP {\bf 0805}   093  (2008)\\
F.G.Scholtz, B.Chakraborty, S.Gangopadhyay and A.Hazra, Phys.Rev.{\bf D71} 085005 (2005)\\
F.G.Scholtz, L.Gouba, A.Hafver and C.M.Rohwer, J.Phys.A:Math.Theor,{\bf 42}  175303(2009)\\
P.A.Horvathy and M.S.Plyushchay,Nucl.Phys \textbf{B714} 269(2005)\\
B.Basu,S.Ghosh and S.Dhar,Europhys.Lett.\textbf{76}  395 (2006)\\
S.Dhar,B.Basu and S.Ghosh  ,Phys.Lett.\textbf{A 371}  406 (2007)\\
\item L. Lucjan, P. Hawrylak, A. Wojs, {\it{Quantum Dots}}, Springer-Verlag, Berlin Heidelberg (1998)\\
  T.Chakraborty, {\it{Quantum Dots A survey of the properties of artificial atoms}},
  Elsevier, The Netherlands (1999).\\
  \item G.Dresselhaus, Phys.Rev. {\bf 100} (1955) 580
 \item  E.I.Rashba, Fiz.Tverd.Tela(Leningrad) {\bf 2} 1224 (1960) , Sov.Phys.Solid State {\bf 2} 1109 (1960);  Yu.A.Bychkov and E.I.Rashba, J.Phys.C {\bf 17} 6039 (1984) \\
 \bibitem{10} A.V.Chaplik and L.I.Magarill, Phys.Rev.Letts. {\bf 96} 126402 (2006) 
 \bibitem{11} T. Chakraborty and P. Pietilainen, Phys. Rev.B {\bf 71} 113305 (2005)
 \bibitem{12} B. Basu and B. Roy, Eur.J. Phys. (accepted) (2009)
 \bibitem{jiang} Jian-zu Zhang, Phys. Lett. B {\bf 584} 204(2004)\\
 \bibitem{gomis}  J. Gomis and T.Mehen, Nucl. Phys. {\bf B 591} 265 (2000), M. Chaichian.et.al. Eur. Phys. J.{\bf C 20}
  767 (2001)
\bibitem{curt}  T. Curtright et.al. Phys. Rev. D {\bf 58} 025002 (1998)
\bibitem{alneir}I.L. Aleiner and V.I. Fal'ko, Phys. Rev. Let. {\bf 87} 256801 (2001)
\bibitem{rod} M. Valin-Rodriguez, Phys.Rev.{\bf B70} 033306 (2004).
\bibitem{15} A. Kijanka and P. Kosinski, Phys. Rev. D {\bf 70}, 1277702 (2004)
\bibitem{16} V.P. Nair and A.P. Polychronakos, Phys. Lett. B{\bf 505} 267 (2001)
\bibitem{17} A. Smailagic and E. Spallucci, Phys. Rev. D {\bf 65}, 107701 (2002)
%\bibitem{18} J.D. Louck et.al., J.Math. Phys. {\bf 14}, 692 (1973)
\bibitem{pr} P.Giri and P. Roy Euro.Phys. J. C {\bf 57 } 835 (2008)
\bibitem{gamboa} J.Gamboa et. al. Mod. Phys. Lett. A {\bf 16} 2075 (2001)

 %\item P.D.Alvarez, J.Gomis, K.Kamimura and M.S.Plyushchay,Phys.Lett {\bf{B659}},906(2008)
 %\item M.Demetrian and D.Kochan,Acta Physica Slovaca \textbf{52,1(2002)}
 % \item P.A.Horvathy,Annal.Phys \textbf{299,128(2002}
 %\item M.A.del Olmo and M.S.Plyushchay,Annals Phys,\textbf{321,2830(2006)}
 %\item P.D.Alvarez,J.Gomis,K.Kamimura and
 %M.S.Plyushchay,Annals.Phys \textbf{322,1556(2007)}
 %\item R.Banerjee and K.Kumar,Phys.Rev. \textbf{D75,045008(2007)}
 %\item A.P.Balachandran,T.R.Govindarajan,C.Molina and
 %P.Teotonio-Sobrinho,JHEP\textbf{0410(2004) 072}.
 %\bibitem{gomis} J. Gomis and T. Mehen Nucl. Phys. B591, 265 (2000)
\end{enumerate}

\end{document}